\documentclass[showkeys,superscriptaddress]{revtex4}
\usepackage{hyperref}
\usepackage{amssymb}
\usepackage{amsmath}
\usepackage{color}

\begin{document}
\title{The motion of color-charged particles \\ 
as a means of testing the non-Abelian dark matter model
}
\author{
Vladimir Dzhunushaliev
}
\email{v.dzhunushaliev@gmail.com}
\affiliation{
Department of Theoretical and Nuclear Physics,  Al-Farabi Kazakh National University, Almaty 050040, Kazakhstan
}
\affiliation{
Institute of Experimental and Theoretical Physics,  Al-Farabi Kazakh National University, Almaty 050040, Kazakhstan
}

\author{
Vladimir Folomeev
}
\email{vfolomeev@mail.ru}
\affiliation{
Institute of Physicotechnical Problems and Material Science of the NAS of the Kyrgyz Republic, 265a, Chui Street, Bishkek 720071,  Kyrgyzstan
}

\author{
Nina Protsenko
}
\email{ninok94kaz@mail.ru}
\affiliation{
Department of Theoretical and Nuclear Physics,  Al-Farabi Kazakh National University, Almaty 050040, Kazakhstan
}

\begin{abstract}
A possibility is discussed for experimental testing of the dark matter model 
supported by a classic non-Abelian SU(3) gauge (Yang-Mills) field. Our approach 
is based on the analysis of the motion of color-charged particles on the 
background of color electric and magnetic fields using the Wong equations. 
Estimating the magnitudes of the color fields near the edge of a galaxy, we 
employ them in obtaining the general analytic solutions to the Wong equations. 
Using the latter, we calculate the magnitude of the extra acceleration of
color-charged particles related to the possible presence of the color fields in the neighbourhood of the Earth.
\end{abstract}



\maketitle

\section{Introduction}
At the present time, it is believed that, beyond ordinary visible  matter, the Universe contains some other, invisible
gravitationally attractive substance. There is a number of observational evidence in favor of its existence~\cite{Roos:2012cc}.
In particular, this applies to the intriguing behavior of the galactic rotation curves. According to the Newtonian theory of gravitation,
the circular velocity $u$ of an object on a stable Keplerian orbit with radius $r$ is $u(r)\varpropto \sqrt{M(r)/r}$,
where $M(r)$ is the mass enclosed within the sphere of radius $r$. Then, in performing observations in the region
beyond the limits of the visible boundary of a galaxy where $M=\text{const.}$, one could expect that the velocity
$u(r) \varpropto 1/\sqrt{r}$. However, current astronomical observations indicate that in the outer regions of most galaxies
$u$ becomes approximately constant. This implies that around the galaxies there exist halos within which
the mass density behaves like $\rho(r) \varpropto 1/r^2$ and the mass $M(r) \varpropto r$.
In addition, the measurements of  peculiar velocities of galaxies in clusters and the effects associated with a gravitational lensing
certainly indicate that these observational consequences also cannot be explained only by the presence of visible matter.

Theoretical modelling of the aforementioned observational effects is usually carried out within the framework of two main directions.
Firstly, it is assumed that the dominating form of matter in galaxies and their clusters is some
invisible substance called  dark matter (DM) (for a general review on the subject, see, e.g., Ref.~\cite{Bertone:2010zza}).
It is commonly believed that in the present Universe DM is about  25\% of the total mass of all forms of matter.
 The true nature of DM is still unknown. It is assumed that it may consist of as yet undiscovered particles.
 This cannot be baryons, since in this case the cosmic microwave background and the large-scale structure of the Universe would look radically different.
 Therefore, as candidates for dark matter particles, various exotic particles which either weakly or not at all interact with ordinary matter and electromagnetic
 radiation are being suggested (axions, sterile neutrinos, gravitinos, weakly interacting massive particles, etc.).
 Moreover, it is assumed that such particles can be confined on scales of galaxies and their clusters.

Second direction of modelling of dark matter is based on the assumption that on galactic scales both the Newtonian and Einsteinian theories of gravity
require a modification~\cite{Bekenstein:1984tv,Capozziello:2012ie}.
This allows the possibility of explaining the aforementioned observational effects without invoking the hypothesis of the presence of DM in galaxies.

In summary, it is seen that the explanation of the observational data requires either introducing new, not yet experimentally discovered  forms of matter or modifying
theories of gravity by themselves. In the present paper, we work within the first approach assuming that galaxies can contain a special type of dark-matter particles modeled
by color fields within the framework of the classical non-Abelian gauge Yang-Mills theory~\cite{litlink12,litlink11,litlink100}. As discussed in those
references, it is possible to construct the gauge field distribution
 which adequately describes the universal rotation curve of spiral galaxies.
 The color DM is described by a special ansatz which enables us to obtain static solutions of the SU(3) Yang-Mills equations.
 Invisibility of such DM is provided by the fact that color particles interact
with ordinary matter and Maxwell's electromagnetic radiation only gravitationally.
 Working within the framework of this model, the aim of the paper is to study the influence
which the presence of such color DM has on the motion of test color-charged particles (monopoles or quarks).
For this purpose, we employ Wong's equations which are solved on the background of color electric and magnetic fields
obtained for external regions of our galaxy (in the neighbourhood of the Earth).

The paper is organized as follows.
In Sec.~\ref{dm_model}, a general description of the non-Abelian dark matter model used in the paper is presented.
In Sec.~\ref{sec_method}, we consider an approach for testing such DM model, within which the magnitudes of color field strengths
are estimated (Sec.~\ref{estim_str_pot}) and Wong's equations are solved analytically (Sec.~\ref{Wong_sol}).
Finally, in Sec.~\ref{concl}, we summarize the obtained results and give some comments concerning the DM model under consideration.

\section{Non-Abelian dark matter model}
\label{dm_model}

In this section we will closely follow Refs.~\cite{litlink12,litlink11,litlink100} where the DM model described by a classical non-Abelian field has been considered.

\subsection{General equations}

It is assumed that a galaxy is embedded in a sphere consisting of SU(3) gauge fields.
The modelling of DM is carried out using the classical SU(3) Yang-Mills equations
\begin{equation}
\label{1-1}
	D_\mu F^{a}_{\mu\nu}=0,
\end{equation}
where
$F^a_{\mu\nu}= \partial_{\mu} A^a_\nu - \partial_{\nu} A^a_\mu + gf^{abc}A^b_\mu A^c_\nu$  is the Yang-Mills field tensor, $A^a_\nu$ is the SU(3) gauge potential, $\mu,\nu = 0,1,2,3$  are spacetime indices,
$g$  is the coupling constant, $f^{abc}$   are the SU(3) structure constants, and $a,b,c= 1, 2, \ldots , 8$ are color indices.

In order to test this DM model experimentally,
one can consider the motion of color-charged particles (monopoles or single quarks) placed in such gauge fields.
The color-charged particles are a non-Abelian generalization of a classical electric charge in Yang-Mills gauge theories.
They are characterized by the color charge $T_a$.

The motion of a test particle with the mass $m$ under the action of external color electric and magnetic fields is described by Wong's equations~\cite{Wong:1970fu}
\begin{eqnarray}
    m c\frac{d^2x^\mu}{ds^2} &=& -g \hbar F_a^{\mu \nu} T_a\frac{dx_\nu}{ds},
\label{1-2} \\
   \frac{dT_a}{ds} &=& -g f_{a b c}
  A_\mu^b \frac{dx^\mu}{ds}   T_c.
\label{1-4}
\end{eqnarray}
The right-hand side of Eq.~\eqref{1-2}
is a generalization of the Lorentz force from Maxwell's electrodynamics to the color fields, and the right-hand side of Eq.~\eqref{1-4} describes the rotation of the vector $T_a$ in the space of color charges.

\subsection{Distribution of color dark matter}

The Yang-Mills equations \eqref{1-1} are solved using the following static ansatz for the classical SU(3) gauge field $A^a_\mu$ \cite{litlink13}:
\begin{eqnarray}
    A^2_0 &=& -2 \frac{z}{gr^2}\chi(r) , \quad
    A^5_0 = 2 \frac{y}{gr^2}\chi(r) , \quad
    A^7_0 = -2 \frac{x}{gr^2}\chi(r) ,
 \label{1-20}\\
    A^2_i &=& 2 \frac{ \epsilon_{3ij} x^j}{gr^2}
    \left[h \left( r \right)+1 \right] ,
    A^5_i = -2 \frac{ \epsilon_{2ij} x^j}{gr^2}\left[h\left( r \right)+1 \right] , \quad
    A^7_i = 2 \frac{ \epsilon_{1ij} x^j}{gr^2}
    \left[h\left( r \right)+1 \right] ,
 \label{1-35}\\
    A^a_i &=& \lambda_{ajk}\left( \epsilon_{ilk} x^j + \epsilon_{ilj} x^k\right) \frac{x^l}{gr^3}v(r) ,
 \quad
    A^a_0 = \frac{1}{2}\left( \lambda_{aij} + \lambda_{aji}\right)x^i x^j\frac{w(r)}{gr^3}.
 \label{1-45}
\end{eqnarray}
Here the components of the gauge field $ A^{2,5,7}_\mu \in \text{SU(2)}\subset \text{SU(3)}$;
 $i,j,k = 1,2,3$ are space indices;
$\epsilon_{ijk}$ is the completely antisymmetric Levi-Civita symbol; $\lambda_{ajk}$ are the Gell-Mann matrices;
$\chi(r), h(r), v(r)$, and $w(r)$ are some unknown functions.
This ansatz is written in cartesian coordinates $x, y, z$ with $r^2=x^2+y^2+z^2$.

Substituting the components \eqref{1-20}-\eqref{1-45} into \eqref{1-1} and setting for simplicity $\chi(r) = h(r) = 0$,
one can obtain the following set of equations for the auxiliary functions $v$ and $w$:
\begin{eqnarray}
  \xi^2 w'' &=& 6w v^2  ,
\label{1-50}\\
  \xi^2v'' &=& v^3-v-v w^2.
  \label{1-51}
\end{eqnarray}
Here the prime denotes differentiation with respect to the dimensionless radius $\xi=r/r_0$, where ${r_0}$  is a constant.
The asymptotic behavior of the functions $v(\xi)$ and $w(\xi)$ at $\xi\gg 1$ is as follows,
\begin{eqnarray}
 v(\xi) &\approx & A \sin(\xi^\alpha +\phi_0) ,
\label{2-10}\\
 w(\xi)&\approx& \pm \left[
 \alpha \xi^\alpha + \frac{\alpha-1}{4}
 \frac{\cos(2 \xi^\alpha + 2 \phi_0)}{\xi^\alpha}\right] ,
\label{2-15}
 \end{eqnarray}
where $\phi_0$ and $\alpha$ are constants, and $A^2 = \alpha(\alpha-1)/3$.

The corresponding DM energy density for the system described by Eqs.~\eqref{1-50} and \eqref{1-51}
can be obtained in the form
\begin{equation}
\label{2-16}\\
 \varepsilon_{\text{DM}}(r) = - F^a_{0i}F^{a0i}+ \frac{1}{4} F^a_{ij} F^{aij} = \frac{1}{g^2 r^4_0} \left[  4\frac{v'^2}{\xi^2}+ \frac{2}{3} \frac{(\xi w'-w)^2}{\xi^4}+2\frac{(v^2-1)^2}{\xi^4}+4\frac{v^2w^2}{\xi^4} \right],
\end{equation}
where the expression in the square brackets is the dimensionless energy density.

Taking into account the asymptotic solutions \eqref{2-10} and \eqref{2-15}, one can show that the gauge field distribution under consideration has an infinite energy,
as a consequence of the asymptotic behavior of the energy density \eqref{2-16} (for more details, see Ref.~\cite{litlink100}).
Consequently, one has to have some cut-off mechanism for such a spatial distribution of the classical gauge fields.

In our opinion, this can be done as follows. As one can see from Eqs.~\eqref{2-10} and \eqref{2-15},
the gauge potentials are oscillating functions whose frequency increases with increasing distance. Then
at large distances from the center the frequency of such oscillations
becomes so large that one has already to take into account quantum fluctuations.
Thus at some distance from the origin the gauge field should undergo the transition from the classical state to the quantum one.
In turn,  the quantum field transits very rapidly to its vacuum expectation value.
Then the distance at which the transition from the  classical state to the quantum one takes place can be regarded as a cut-off radius
 up to which the solutions of Eqs.~\eqref{1-50} and \eqref{1-51} remain valid.
Also, it is very important here to note that the gauge field in the vacuum state must be described
in a nonperturbative  manner (see below in Sec.~\ref{transition}).

\subsection{Invisibility of  color fields}

Let us now say a few words about the main feature of any dark matter -- its invisibility.
Within the framework of the DM model under consideration, this is achieved in a very simple way:
the SU(3) color matter (dark matter in the context of the present paper)
is invisible because color gauge fields interact only with color-charged particles;
but at the present time particles possessing a SU(3) color charge have not still been registered experimentally.

In principle, as a candidate for such particles, one may consider SU(3) monopoles. 
For such a case, one can write down the SU(3) Lagrangian
\begin{equation}
	L_{\text{QCD}} = -\frac{1}{2} \text{Tr} \left(F^a_{\mu\nu}F^{a \mu \nu}\right)+ \sum\limits_{k}^{n_f}  \overline{q}_k \left( i \gamma^\mu D_\mu -m_k \right)q_k
\label{2-17}
\end{equation}
describing monopoles interacting with matter in the form of quarks. Here
\begin{equation}
	D_\mu q =\left(\partial_\mu-igA_\mu \right)q \quad \text{with} \quad
A_\mu = \sum\limits_{b=1}^{8} A^b_\mu \frac{\lambda^b}{2}
	\label{2-18}
	\end{equation}
(the gauge potential $ A_\mu $ is given here in the matrix form). From the term   $i g A_\mu q$ in Eq.~\eqref{2-18},
one can see that the SU(3) color field interacts with quarks only. But free quarks are not observable in nature.
All other forms of matter are colorless, including baryonic matter (as a consequence of the confinement of quarks in hadrons) and photons.
Therefore color DM being considered here does not interact with them directly and can be observable only through its interaction with a gravitational field.
It is interesting that in this respect the problem of dark matter in astrophysics is related to the problem of confinement in high energy physics.

\subsection{Transition from the classical phase to the quantum one}
\label{transition}

As emphasised earlier,  in describing DM we use a classical SU(3) gauge field which undergoes the transition from the classical state to the quantum one
at some distance from the center of a galaxy.
Without such a transition, the energy of the sphere filled with this field would be infinite. We assume that by taking into account
nonperturbative quantum effects the energy of such a field configuration can be made finite.
Here we consider a possibility, in principle, of introducing a cut-off mechanism for  the spatial distribution of the classical gauge fields
at some distance from the center of a galaxy.

To do this, let us employ the Heisenberg uncertainty principle according to which
\begin{equation}
	\frac{1}{c} \; \Delta F^a_{ti} \; \Delta A^{a i} \; \Delta V \approx \hbar.
\label{4-10}
\end{equation}
Here $\Delta F^a_{ti}$ is a quantum fluctuation of the color electric field $F^a_{ti}$; $\Delta A^{a i}$ is a quantum fluctuation of the color potential $A^{a i}$; $\Delta V$
is the volume where the quantum fluctuations $\Delta F^a_{ti}$ and $\Delta A^{a i}$ occur.

Using the ansatz \eqref{1-20}-\eqref{1-45}, we have the following components of the gauge potential written in  spherical coordinates:
\begin{eqnarray}
	A^a_t &=& \frac{1}{g r}\Big\{
		w(r) \sin^2 \theta \sin(2 \varphi);   \quad
		-2 \chi(r) \cos \theta ;   \quad
		w(r) \sin^2 \theta \cos (2 \varphi);   \quad
		w(r) \sin(2 \theta) \cos \varphi ;
	\nonumber \\
	&&
			2 \chi(r) \sin \theta \sin \varphi;  \quad
		w(r) \sin(2 \theta) \sin \varphi;  \quad
		- 2\chi(r) \sin\theta\cos \varphi;  \quad
		-w(r) \frac{1 + 3 \cos(2 \theta)}{2 \sqrt{3}}
	\Big\};
\label{a1-80}\\	
	A^a_r &=& 0;
\label{a1-90}\\
	A^a_\theta &=& \frac{1}{g}\Big\{
	-2  v(r)\sin \theta\,\cos \left(2\varphi\right); \quad
	0; \quad
	2  v(r)\sin \theta \sin\left( 2\varphi \right) ;  \quad
	2  v(r)\sin \varphi \, \cos \theta  ;  \quad
	2 \left[ 1 + h(r) \right] \cos \varphi ;
	\nonumber \\
	&&
		-2 v(r) \cos \theta \cos \varphi ;  \quad
	2 \left[ 1 + h(r) \right] \sin \varphi ;  \quad
	0
	\Big\};
\label{a1-100}\\
	A^a_\varphi &=& \frac{1}{g}\Big\{
		v(r)\sin \theta \sin(2 \theta) \sin(2 \varphi) ;   \quad
		-2 [1 + h(r)] \sin^2 \theta ;   \quad
		2 v(r) \sin^2 \theta \cos \theta \cos (2 \varphi ) ;  \quad
		 v(r)  \cos \varphi \left(\sin (3\theta)-\sin\theta\right);
	\nonumber \\
	&&
		- [1 + h(r)] \sin (2\theta) \sin \varphi ;   \quad
	2 v(r) \sin \theta \sin \varphi \cos (2 \theta);   \quad
	 [1 + h(r)] \sin (2\theta)\cos \varphi ;   \quad
	\sqrt{3} v(r) \sin \theta \sin (2 \theta)
	\Big\}.
\label{a1-110}
\end{eqnarray}

Using these components and taking into account that in our case  $h(r)=\chi(r)=0$, one can find that there are three non-zero components of the color
electromagnetic field tensor  $F^2_{t \theta}, F^5_{t \varphi}$, and $F^7_{t \varphi}$ that can appear  in the left-hand side of the relation~\eqref{4-10},
and all of them $\varpropto v w/(g r)$. For our purpose, we can either employ all three components or take any one of them.
In the latter case, calculations are more simple and they can give the required rough estimate.
Let us therefore use in \eqref{4-10} the following component:
\begin{equation}
	E^2_\theta=F^2_{t \theta} = - \frac{2}{g} \sin \theta \frac{v w}{r} .
\label{3-200}
\end{equation}
Introduce the physical component $\tilde F^2_{t \theta}$,
\begin{equation}
	\left| \tilde F^2_{t \theta} \right| =
	\sqrt{\left|F^2_{t \theta} F^{2 t \theta}\right|} = \frac{2}{g} \sin \theta \frac{v w}{r^2} .
\label{3-300}
\end{equation}
Then, apart from a numerical factor, the fluctuations of the SU(3) electric field  are
\begin{equation}
	\Delta \tilde F^2_{t \theta} \approx \frac{1}{g} \frac{1}{r^2} \left(
		\Delta v \; w + v \; \Delta w
	\right).
\label{4-20}
\end{equation}

In turn, from Eqs.~\eqref{a1-80}-\eqref{a1-110}, we have
\begin{equation}
	A^2_\theta = 0, \quad
	A^{1,3,4,6}_\theta \propto \frac{1}{g} v.
\label{3-85}
\end{equation}
Introducing the physical components 
\begin{equation}
	\left| \tilde A^{1,3,4,6}_\theta \right| =
	\sqrt{A^{1,3,4,6}_\theta A^{1,3,4,6\,\, \theta}} \approx
	\frac{1}{g}  \frac{v}{r},
\label{3-90}
\end{equation}
we assume that
\begin{equation}
	\Delta \tilde A^2_\theta \approx \Delta \tilde A^{1,3,4,6}_\theta \approx
	\frac{1}{g}  \frac{\Delta v}{r} .
\label{4-30}
\end{equation}

Next, the period of spatial oscillations at $r \gg r_0$ can be defined as follows,
\begin{equation}
	\left( \xi + \lambda \right)^\alpha - \xi^\alpha \approx
	\alpha \frac{\lambda}{\xi^{1 - \alpha}} = 2\pi;
	\quad \xi = \frac{r}{r_0}.
\label{3-120}
\end{equation}
Suppose that the distance at which the classical SU(3) color field becomes quantum one is defined as the radius where magnitudes of the
quantum fluctuations of the field enclosed in the volume
\begin{equation}
	\Delta V = 4\pi r^2 \Delta r \quad \text{with} \quad \frac{\Delta r}{r_0} \approx \lambda
	\approx \frac{1}{\alpha} \frac{2 \pi}{\xi^{\alpha - 1}}
\label{4-40}
\end{equation}
 become comparable with magnitudes of this classical field. That is, at the transition distance, we assume that
\begin{equation}
	\Delta v \approx v, \quad
	\Delta w \approx w.
\label{3-140}
\end{equation}
Substituting  Eqs.~\eqref{2-10}, \eqref{2-15}, \eqref{4-20},  \eqref{4-30}, \eqref{4-40},  and \eqref{3-140} into~\eqref{4-10}, we obtain
\begin{equation}
	\left( \frac{g'}{A} \right)^2 \approx 2 \pi,
\label{3-150}
\end{equation}
where $g' = \sqrt{\hbar c/4\pi} g$ is the dimensionless coupling constant, similar to the fine structure constant in quantum electrodynamics $\alpha = e^2/\hbar c$.
In quantum chromodynamics,
$\beta = 1/{g'}^2 \gtrsim 1$. If one chooses ${g'} \approx 1$ and  $A \approx 0.4$
(this value follows from the numerical computations of Refs.~\cite{litlink12,litlink11,litlink100}), then
\begin{equation}
	\left( \frac{g'}{A} \right)^2 \approx 6.25,
\label{3-160}
\end{equation}
that is comparable with $2 \pi \approx 6.28$.

Thus we have shown that if the condition \eqref{3-150} holds at some distance from the center, then there occurs
the transition from the classical phase to the quantum one.
Unfortunately, the obtained rough estimate does not enable us to calculate the radius at which such a transition takes place.
For finding this radius, it is necessary to have \emph{nonperturbative} quantization methods which are absent at the moment.

\section{An approach for testing the non-Abelian dark matter model}
\label{sec_method}

\subsection{Estimating the field strength and gauge field potentials}
\label{estim_str_pot}

We propose here an approach which permits us to test the DM model described above
by studying the motion of a color-charged particle (monopole or single quark) under the action of color electromagnetic fields.
For this purpose, we will use the Wong equations \eqref{1-2} and \eqref{1-4}.
To simplify them, we restrict ourselves to considering a trajectory of a particle moving in the equatorial plane (i.e., when $\theta = \pi / 2$)
at the fixed distance from the center $r = \text{const}.$
Also, since the dimension of the experimental set-up is assumed to be much less than the radius of a galaxy,
one can set the angle variable $\varphi \approx 0$. In this case the potential and the field strength are especially simple.
Taking all this into account, let us write down the  non-zero components of the color electromagnetic field strength and of the gauge potential:
\begin{eqnarray}
   &&E^2_\theta = E^5_\varphi  =  - \frac{2v w}{g r},
\quad
  E^3_r = \sqrt 3 E^8_r =- \frac{r w' - w}{g r^2},
\label{2-30}\\
  &&- H^1_\varphi = H^4_\theta = \frac{2 v'}{g} ,
\quad
	H^7_r = \frac{2}{g r^2} \left( v^2 - 1 \right) ,
\label{2-33}\\
	&&A^1_\theta = A^4_\varphi = -\frac{2 v}{g} ,
\quad
	A^3_t = \sqrt 3 A^8_t = \frac{w}{g r}
\label{2-34},
\end{eqnarray}
where the functions $v$ and $w$ appearing here are the asymptotic solutions given by Eqs.~\eqref{2-10} and \eqref{2-15}.

For a rough estimate of the magnitudes of the color fields, let us employ Newton's law of gravitation which gives the following relation
for a test particle located near the edge of a galaxy and rotating around its center with the circular velocity $u$,
\begin{equation}
  \frac{u^2}{r_g} = \gamma \frac{M}{r^2_g}.
\label{2-35}
\end{equation}
Here $\gamma$ is the Newtonian gravitational constant;
$M = M_v + M_{\text{DM}}$ is the total mass of the galaxy, including the masses of  visible, $M_v$, and dark, $M_{\text{DM}}$, components;
$r_g$ is the radius of the galaxy.
From this relation, one can find  the mass of DM as
\begin{equation}
  M_{\text{DM}} = \frac{u^2}{\gamma} r_g - M_v.
\label{2-45}
\end{equation}

The radial distribution of the energy density of the color electromagnetic field describing  DM is
\begin{eqnarray}
 \varepsilon_{\text{DM}} = - \frac{1}{2}\left(
 	E^a_i E^{a i} + H^a_i H^{a i}  \right)&\approx& - \frac{1}{2}\left(
 	E^2_\theta E^{2 \theta} + E^5_\varphi E^{5 \varphi} +
 	E^3_r E^{3 r}+E^8_r E^{8 r} + H^4_\theta H^{4 \theta} +
 	H^1_\varphi H^{1 \varphi}	\right) \nonumber \\
 &\approx &	\frac{1}{ g^2 r^4}\left[\frac{2}{3}\left(r w^\prime-w\right)^2+4 r^2 v^{\prime 2}+4 v^2 w^2
 \right].
\label{2-50}
\end{eqnarray}
Here $E^a_i = F^a_{ti}$ is the chromoelectric field and
$H^a_i = \frac{1}{2} \sqrt{-g} \epsilon_{ijk} F^{a jk}$ is the chromomagnetic field.
The expression  \eqref{2-50} corresponds to the energy density \eqref{2-16}, where the asymptotic solutions
 \eqref{2-10} and \eqref{2-15} have been used and only the terms giving leading contributions have been kept.

On the other hand, the DM energy density occurs in the expression
\begin{equation}
  M_{\text{DM}} \approx \frac{4}{3}  \pi r^3_g \frac{\varepsilon_{\text{DM}}}{c^2}
\label{2-55}
\end{equation}
(for simplicity, we assume here that the electric and magnetic fields are distributed  homogeneously).
Comparing the expressions \eqref{2-45} and \eqref{2-55} and neglecting the mass of the visible component $M_v$ compared with $M_{\text{DM}}$,
we can get the following rough estimates for the field strengths:
\begin{equation}
  \tilde E^2_\theta \approx \tilde E^5_\varphi  \approx
 	\tilde E^3_r \approx \tilde E^8_r  \approx  \tilde H^4_\theta  \approx
 	\tilde H^1_\varphi  \approx B \quad \text{with} \quad B=\sqrt{\frac{3}{4\pi\gamma}}c\frac{u}{r_g} ,
\label{2-60}
\end{equation}
where the physical components of the fields are
\begin{align}
\label{phys_E}
\begin{split}
&\tilde E^2_\theta=\sqrt{E^2_\theta E^{2 \theta}}=\frac{2 v w}{g r^2},\quad
 \tilde E^5_\varphi=\sqrt{E^5_\varphi E^{5 \varphi}}=\frac{2 v w}{g r^2},\quad
 \tilde E^3_r=\sqrt{E^3_r E^{3 r}}=\frac{r w' - w}{g r^2}, \\
&\tilde E^8_r=\sqrt{E^8_r E^{8 r}}=\frac{1}{\sqrt{3}}\frac{r w' - w}{g r^2}, \quad
  \tilde H^4_\theta=\sqrt{H^4_\theta H^{4 \theta}}=\frac{2 v'}{g r}, \quad
  \tilde H^1_\varphi=\sqrt{H^1_\varphi H^{1 \varphi}}=\frac{2 v'}{g r}.
\end{split}
\end{align}

In turn, using Eq.~\eqref{2-34}, one can introduce the following physical components for the potential:
\begin{align}
\label{2-80}
\begin{split}
 &\tilde A^1_\theta=\sqrt{A^1_\theta A^{1 \theta}}=\frac{2 v}{g r}, \quad
  \tilde A^4_\varphi=\sqrt{A^4_\varphi A^{4 \varphi}} = \frac{2 v}{g r} , \\
&\tilde A^{3}_t=\sqrt{A^{3}_t A^{3 \; t}} =\frac{w}{g r}, \quad \tilde A^{8}_t=\sqrt{A^{8}_t A^{8 \; t}} =\frac{1}{\sqrt{3}}\frac{w}{g r}.
\end{split}
\end{align}

The obtained estimates will be used below when considering the motion of test particles under the action of the given color fields.

\subsection{Solving Wong's equations}
\label{Wong_sol}

As mentioned earlier,  the dimension of the experimental set-up for studying the motion of color-charged particles is assumed to be much less than the radius of a galaxy.
Assuming also that the velocities of test particles are small compared to $c$,
 it is sufficient to consider the nonrelativistic limit of the Wong equations. In this case,
 setting $ds\approx cdt$, we have from Eqs.~\eqref{1-2} and \eqref{1-4}
\begin{eqnarray}
   m\frac{d^2x^i}{dt^2} &=& -g \hbar c F^{i t}_{a} T_a,
\label{1-3} \\
   \frac{dT_a}{dt} &=& -g c f_{a b c}  A_t^b  T_c.
\label{1-31}
\end{eqnarray}
As before, here  $i=1,2,3$ is a space index. As pointed out in section \ref{estim_str_pot},
we consider the case where the chromoelectric and chromomagnetic field strengths are of the same order of magnitude.
This allowed us to neglect the spatial components of the four-velocity in the above equations. Correspondingly, Eq.~\eqref{1-3}
contains now only the components $F^{i t}_{a}$ describing the chromoelectric field, but not the terms with the chromomagnetic field.

For the sake of simplicity,
it is convenient to solve Eqs.~\eqref {1-3} and \eqref {1-31} in cartesian coordinates $x,y,z$, where the coordinate
$x$ is directed along the radius $r$, the coordinates $y$ and $z$ -- along the angle variables $\theta$ and $\varphi$, respectively.
The resulting set of equations describing the motion of a test particle with the mass $m$ and the dynamics of the color charge vector $T_a$ is
\begin{eqnarray}
 &&m \frac{d^2 x}{dt^2}  = -g \hbar c  E^3_{r}\left(T_3+\frac{1}{\sqrt{3}}T_8\right),
\quad
  m \frac{d^2 y}{dt^2}  = -g \hbar c  E^2_{\theta}T_2,
\quad
  m \frac{d^2 z}{dt^2}  = -g \hbar c  E^5_{\varphi}T_5,
\label{3-15}\\
  &&\frac{dT_1}{dt} = g c A_t^3 T_2, \quad \frac{dT_2}{dt} = -g c A_t^3 T_1,
\quad
  \frac{dT_4}{dt} = g c A_t^3 T_5, \quad \frac{dT_5}{dt} = -g c A_t^3 T_4,
  \quad
  \frac{d T_{3,6,7,8}}{d t}=0.
  \label{3-25}
\end{eqnarray}
The numerical values of the components  $E^3_{r}, E^2_{\theta}, E^5_{\varphi}$, and $ A_t^3$ appearing here are taken from the estimates
\eqref{2-60} and \eqref{2-80}.

Eqs.~\eqref{3-15} and \eqref{3-25} have the following general solution:
\begin{align}
& x=x_0+v_{0 x}t-\frac{1}{2}\frac{g\hbar c}{m}E^3_{r}\left(T_3+\frac{1}{\sqrt{3}}T_8\right)t^2, \quad y=y_0+v_{0 y}t+\frac{\hbar }{m g c}\frac{E^2_{\theta}}{\left(A_t^3\right)^2} T_2(t),
\quad z=z_0+v_{0 z}t+\frac{\hbar }{m g c}\frac{E^5_{\varphi}}{\left(A_t^3\right)^2} T_5(t),
\label{3-35}\\
\begin{split}
& T_1(t)=T_1(0) \cos \omega t+T_2(0)\sin \omega t, \quad
T_2(t)=T_2(0) \cos \omega t-T_1(0)\sin \omega t, \\
&T_4(t)=T_4(0) \cos \omega t+T_5(0)\sin \omega t, \quad
T_5(t)=T_5(0) \cos \omega t-T_4(0)\sin \omega t,\\
&T_3=T_6=T_7=T_8=\text{const.},
\label{3-45}
\end{split}
\end{align}
where $x_0, y_0, z_0, v_{0 x}, v_{0 y}, v_{0 z}, T_1(0), T_2(0), T_4(0), T_5(0)$ are integration constants.
These solutions describe oscillations of the field with frequency $\omega=g c A_t^3$ plus the transient motion with the given initial velocities
$v_{0 x}, v_{0 y}, v_{0 z}$ directed along the axes $x, y, z$, respectively.

Let us now estimate the acceleration of a color charge. This can be done using Eq.~\eqref{3-35}, from which we have:
\begin{enumerate}
\itemsep=-0.2pt
\item[(1)]
The $x$-component of the acceleration is
\begin{equation}
a_x=\frac{g\hbar c}{m} E^3_{r}\left(T_3+\frac{1}{\sqrt{3}}T_8\right).
\label{3-50}
\end{equation}
Substituting into this expression the numerical estimate for $E^3_{r}=-\tilde E^3_{r}\approx -B$ [see Eqs.~\eqref{2-30}, \eqref{2-60}, and \eqref{phys_E}]
and taking into account that $g= \sqrt{4\pi/(\hbar c)}g'$, we have
\begin{equation}
a_x\approx \sqrt{\frac{3\hbar c^3}{\gamma}}\frac{g' u}{m r_g}\left(T_3+\frac{1}{\sqrt{3}}T_8\right).
\label{3-51}
\end{equation}

If for a test particle we choose, say, the 't~Hooft-Polyakov monopole with the mass $m\approx 10^{-8} \text{g}$, then, taking into account that
the radius of our galaxy is $r_g \approx 10^{23}$~cm and the velocity of DM particles at the edge of the galaxy  $u\approx 2.5 \times 10^7 \text{cm sec}^{-1}$,
we have
\begin{equation}
a_x\approx 0.03\, g' \left(T_3+\frac{1}{\sqrt{3}}T_8\right) \text{cm sec}^{-2}.
\label{3-52}
\end{equation}
By choosing  $g'\sim 1$ and assuming that the expression in the brackets is also $\sim 1$, one sees that the magnitude of the extra acceleration related to the presence
of the color fields in the neighbourhood of the Earth is of the order of $3\times 10^{-3}\%$ of the free-fall acceleration on the Earth.
Evidently, for monopoles with smaller masses there will be even larger extra accelerations.

\item[(2)] To estimate the accelerations along the coordinates $y$ and $z$, it is necessary to calculate the frequency of oscillations
 $\omega$ from \eqref{3-45}. To do this, let us assume that the numerical values of the components
  $E^2_{\theta}, E^5_{\varphi}$ appearing in Eq.~\eqref{3-35} are approximately equal to the numerical values of the corresponding physical components
 $\tilde E^2_{\theta}, \tilde E^5_{\varphi}$ from \eqref{2-60} and \eqref{phys_E}. That is, we assume that
  $E^2_{\theta}= E^5_{\varphi}\approx B$. Also, taking into account Eqs.~\eqref{2-34} and \eqref{2-80}, one can assume as a rough estimate that the component of the
  potential  $A_t^3\approx r_g B$. Then the resulting frequency is
 $$
 \omega=g c A_t^3=g' \sqrt{\frac{3 c^3}{\gamma \hbar}}u\approx g'\times 10^{40} \text{sec}^{-1}.
 $$
It is seen from this expression that the color field exhibits extremely high-frequency oscillations.
Correspondingly, there will be a large number of oscillations during the time of the experiment,
and one can therefore average the functions $T_2$ and $T_5$ from \eqref{3-45} over these oscillations. The result is $T_2= T_5=0$,
and correspondingly Eq.~\eqref{3-35} yields
 $$
  y=y_0+v_{0 y}t, \quad  z=z_0+v_{0 z}t,
 $$
 i.e., a uniform motion of a test particle along the coordinates $y, z$ with $a_y=a_z=0$.

\end{enumerate}

\section{Conclusion and final remarks}
\label{concl}

The nature of dark matter is one of key questions of modern cosmology and astrophysics. The most popular hypothesis is the assumption that DM
consists of some exotic particles which either weakly or not at all interact with ordinary matter and electromagnetic radiation.
This has the result that their experimental  detection runs into great difficulty.
Various experimental searches have been made to find such particles
(see, e.g., Ref.~\cite{Strigari:2013iaa} where direct and indirect methods are reviewed),
but unfortunately they have so far yielded a negative result.

In the present paper we suggest a way for testing the DM model supported by color electric and magnetic fields described within the framework of the classical SU(3)
Yang-Mills theory. Apart from the gravitational interaction with other forms of matter, such fields can interact directly only with color-charged particles
like monopoles and single quarks. This has the result that, when studying the motion of color-charged particles
in a laboratory on the Earth, there will appear an extra acceleration provided by the color fields.

To calculate the magnitude of such extra acceleration, we have used the known Wong equations describing the classical motion of non-Abelian particles
which are acted upon by the background color fields. When solving this problem, one needs to find the magnitudes of the color field strengths in the neighbourhood of the Earth.
In order to make a rough estimate, we have considered the motion of test particles under the action of the DM gravitational field near the edge of a galaxy.
Using the obtained estimates for the field strengths, we have found general analytic solutions to the nonrelativistic Wong equations which enabled us to calculate the magnitude
of the extra acceleration. It was shown that it is inversely proportional to the mass of a test particle [see Eq.~\eqref{3-51}]. For a grand-unified-theory monopole the extra acceleration is
of the order of a few thousandth of a percent of the free-fall acceleration on the Earth.
In principle, such magnitudes can be registered experimentally.

Let us now give a word about strong and weak sides of the non-Abelian dark matter model used here. Notice, first of all, that practically all known dark matter models
suggest the existence of as yet not discovered forms of matter.
This is obviously a weak side of such models. That is why the fact that within the model considered here DM is described by the well-known
non-Abelian  SU(3) gauge field is a strong side of such a model.

As a weak side of the DM model discussed here one can consider the assumption itself on the possibility of the presence of a classical non-Abelian field on galactic scales.
It can be said here that there are no strong objections to the existence of such fields.
Thus, for example, there exist classical Abelian U(1) gauge electric and magnetic fields. Hence the assumption of the existence of classical non-Abelian fields
is, in principle, similar to the assumption of the existence of classical Abelian fields.

In actuality, the principle difference is in an asymptotic behavior of the fields: the Abelian fields have the Coulomb behavior and the non-Abelian fields~--
the non-Coulomb one (they decrease slower than $1/r^{2}$). In this connection, at a qualitative level, we have discussed a possible solution of this problem (see in Sec.~\ref{transition}).

Finally, notice that the motion of color-charged particles studied here can be regarded as a problem of primarily academic interest. Indeed,
from the experimental point of view the most evident weak side of the non-Abelian DM model is the actual absence of test particles:
monopoles have not yet been discovered; free quarks do not exist (at low energies/temperatures). Nevertheless, experimental searches for magnetic monopoles continue \cite{Patrizii:2015uea} and perhaps they
will be discovered in the future.

\section*{Acknowledgments}

The authors gratefully acknowledge support provided by a Grant IRN: BR05236494 
of the Ministry of Education and Science of the
Republic of Kazakhstan.

\end{document}